# On the influence of uncertainties in scattering potentials on quantitative analysis using keV ions

Barbara Bruckner[a,b,*], Tomas Strapko[a,1], Mauricio A. Sortica[a], Peter Bauer[b] and Daniel Primetzhofer[a]

[a] Department of Physics and Astronomy, Uppsala University, Box 516, S-751 20 Uppsala, Sweden

[b] Johannes-Kepler Universität Linz, IEP-AOP, Altenbergerstraße 69, A-4040 Linz, Austria

*barbara.bruckner@physics.uu.se

[1] Current address: Department of Physical Engineering and Nanotechnology, Brno University of Technology, Technická 2896/2, 616 69 Brno-Královo Pole, Czech Republic

**Abstract**

Experimental spectra from medium energy ion scattering were compared to Monte-Carlo simulations (employing the TRBS code) to obtain information on the scattering potential. The impact of uncertainties in the interatomic potential on quantification of sample properties such as thickness, composition or electronic stopping was investigated for different scattering geometries: backscattering and transmission. For backscattered He ions with tens of keV primary energy the scattering potential was found to overestimate the multiple scattering background in the energy spectra resulting in an uncertainty of < 3 % in quantitative analysis. Light ions transmitted through a sample for equivalent path length in the medium are only affected minorly by changes in the scattering potential. This effect becomes more distinct for heavier primary ions.

# 1. Introduction

Ion beam analysis is a powerful family of techniques, which has been successfully applied for materials characterization for several decades [1–3]. For analytical methods based on ion scattering, when employing energies on the order of MeV/u, high accuracy and an almost complete absence of matrix effects can be expected [4]. At the same time ion beams have been equally successfully applied for materials modification in both irradiation studies and for ion implantation [5,6].

The simplicity of ion-solid interaction at sufficiently high energies is e.g. reflected by the accuracy achieved by the classical Rutherford cross section describing the nuclear scattering process [7]. This accuracy forms the basis for the straightforward quantification of sample composition, as concentrations can be trivially linked to the scattering yield. The interaction with the electronic system of the target can be described as adiabatic. This simplification allows for good quantitative predictions of the inelastic energy loss of the intruding projectiles [8], and provides accurate ion ranges and depth perception.

In course of the ongoing miniaturization of electronics and sensors new methods for profiling thin films with enhanced depth resolution have become more important. The obtained sub-nanometer resolution achievable in Medium-Energy Ion scattering (MEIS) [9] or Low-Energy Ion scattering (LEIS) [10] is, however, coupled to more complex ion-solid interaction processes at the relevant energies. In particular, the scattering probabilities of both nuclei and electrons are more strongly affected by the surrounding electrons. At the same time, the reduced speed of the primary particles enables a more direct response of the system (as electron screening).

This non-adiabatic character leads for example to an increased importance of charge exchange processes [11] which have been found to show complex behavior affecting quantification [12,13]. Consequently, also the energy dissipation of primary ions shows deviations from expectations by quasi-static predictions [14]. Thus, an accurate description of the dynamic processes induced by the comparably slow intruder into the local electron structure is necessary [15].

Also, the interaction of projectiles and target nuclei is changing for decreasing energies, i.e. the Rutherford cross section is no longer providing an accurate description. The reason for this observation is found in the increased relevance of screening of the nuclear charge by electrons. This phenomenon can be observed even at elevated energies for large interaction distances, which in the present energy regime, however, correspond to large angle collisions. Andersen provided an analytical correction to the Rutherford cross section which is reliable in an extended energy range towards lower energies [16]. Corrections with a more general applicability can be obtained by a modified description of the interaction potential in form of a screened Coulomb potentials such as the Thomas-Fermi-Moliere (TFM) potential [17] or the Universal-potential (ZBL) [18]. In equation (1) the typical expression for a screened Coulomb potential $V_c$ is given. Here $r$ denotes the distance of the interaction partners with $a$, the characteristic screening length, for which different models were proposed, e.g. the Firsov screening length $a_F$ [19].

$$V(r) = V_c(r) * \Phi\left(\frac{r}{a}\right)$$

However, soon after its introduction, this concept has been subject of suggested, typically linear, corrections to the commonly employed screening length $a$ in these potentials, e.g. by O'Connor and Biersack [20]. This development was triggered by the increasing importance of ion-solid interaction at lower energies to further optimize description at large interaction distances. These correction factors have been determined in a number of studies [21–24] with results depending on energy, and experimental geometry. Also independent models of, however, similar character have been proposed [25,26]. In parallel, it is obvious, that the descriptions given above represent only an approximation, as for example both charge state and the excitation state of the primary ion in the solid might be subject to fluctuations depending on interaction distance, which will lead to different potentials applicable [27]. When deducing potential strength from experiments, this fact complicates obtaining a global picture. Obviously, the specific impact parameters probed in an experiment will yield a description of the potential for the relevant interaction distances. The deduced model, assuming a linear correction factor, might be inappropriate in a more global perspective in case the correction is actually non-linear. Note, that impact parameters and the associated interaction distances of relevance, can even for identical ion energies differ significantly in different experimental approaches: In ion implantation, all kind of trajectories and consequently all impact parameters, contribute necessarily to the final modification of the target material. In ion-beam based analysis, in different experimental approaches such as backscattering and transmission experiments designed to measure thickness, very different impact parameter regimes can be expected to be of interest [28,29]. Intrinsically, as not all primary ions are contributing to the relevant signal, both approaches will to certain extend be selective to specific trajectories. Note, that this selection can be made via the scattering geometry as well as via the spectral features relevant for the analysis. These facts imply, that even if a certain choice of the potential has been proven correct in one experimental approach and for a specific ion energy, the model will not necessarily accurately describe a different experiment at similar energies.

In the present study we have systematically investigated, how the specific choice of the scattering potential influences the shape of spectra obtained upon interaction of ions with energies of several tens of keV with materials. Different geometries and target systems have been modelled with computer simulations, and – where applicable – compared to experimental spectra. From the results, we estimate how uncertainties on the specific potential can enter quantities deduced from the specific experiment, such as composition, layer thickness or material properties such as the specific electronic energy loss. We also attempt to link the results obtained to the regime of Low-energy Ion Scattering, in which the influence of screening, and thus its uncertainty has an even stronger impact on quantification.

## 2. Experiment and simulations

Experiments were conducted in the time-of-flight (ToF) MEIS setup located at the University of Uppsala [30]. Here, a 350 kV Danfysik ion implanter is used to produce atomic or molecular ion beams with typical energies between 20 keV and 350 keV. An angularly rotatable, position sensitive detector based on micro-channel plates is employed to detect backscattered particles in a total solid angle of 0.13 sr. In the experimental spectra shown in this contribution, only projectiles backscattered within $\vartheta = 155° \pm 2°$ are considered for evaluation and the obtained ToF-spectra typically have a time resolution of $1 - 2$ ns [31].

The increasing scattering cross sections with decreasing energy of the primary ions generally yield significant contributions from multiple small angle scattering (MS) as well as plural scattering by comparably large angles (PS) which are commonly observable in MEIS and LEIS spectra. Therefore, for analyzing the impact of the scattering potential on spectrum shape and for simulating the experimental spectra we performed Monte-Carlo (MC) simulations. Specifically, we use the TRBS code (TRim for BackScattering) [32], which permits to calculate energy spectra of transmitted and backscattered projectiles as well as depth profiles of implanted particles.

In this program, nuclear scattering is modelled by either the TFM or the ZBL potential using the according screening models. Additionally, for both models a linear screening correction $c_a$ can be applied. For electronic stopping $S_e$ SRIM [18] values are used for which a linear correction can be applied.

To decrease computational time the TRBS code only considers scattering angles $\vartheta > \vartheta_{\min}$ explicitly, with a global description of multiple small angle scattering (see [32] for details), where the cutoff angle $\vartheta_{\min}$ is an adjustable input parameter. Typically, values of $\vartheta_{\min} < 3°$ are used, and the program provides the number of average collisions explicitly experienced by backscattered stopped and transmitted ions respectively.

In order to compare energy spectra obtained from the experiment and simulation the experimental resolution and energy loss straggling has to be considered. Therefore, simulated data are convoluted with a Gaussian of appropriate width.

Two different samples were investigated in the experiments performed for this contribution: a thin HfN film on carbon as well as a TiN film with a W interlayer on a Si substrate. Both samples were characterized by Rutherford backscattering spectrometry (RBS) with a beam of 2 MeV He$^+$ ions to obtain stoichiometry and thickness in terms of atoms/cm$^2$. The composition of TiN was additionally measured in a ToF elastic recoil detection setup with a primary beam of 36 MeV $^{127}$I$^{8+}$. Both experiments were conducted at the tandem accelerator at the Ångström laboratory (Uppsala University). For the TiN we obtained a thickness of 175 Å (assuming a bulk density of 5.43 g/cm$^3$) with Ti:N = 1:1 and a thickness of 13 Å for the W interlayer to the substrate. The HfN sample is 132 Å thick (assuming a bulk density of 13.83 g/cm$^3$) with a stoichiometry of Hf:N = 1:1.27 and 2 % Ar as well as 2.5 % O and 2.5 % Zr.

## 3. Results and discussion

**3.1 Backscattering measurements**

3.1.1 Surface layer analysis

In analytical techniques based on ion scattering, a spectrum obtained in backscattering geometry typically allows to extract information on sample composition, i.e. abundance of certain species (via the signal intensity) and thickness of layers (via the width of characteristic features). In Fig. 1 we show an energy converted ToF-MEIS spectrum of 25 keV He$^+$ scattered from HfN (black open squares). The high energy edge, indicated in the figure as $kE_0$, corresponds to He scattered from Hf in the first monolayer, where $k$ is the kinematic factor. As in typical RBS spectra the width of the Hf peak $\Delta E$ indicated in the figure depends on the film thickness $n\Delta x$. In a single scattering model as commonly used in RBS analysis, the stopping cross section factor $[\varepsilon]$ can be employed. This factor considers energy loss on the way in and out of the sample and allows evaluation of the thickness from: $\Delta E = [\varepsilon] \cdot n\Delta x$ [4]. Different from the regime of MeV energies, the high intensity of the multiple scattering background complicates this analysis as the width of the spectrum is more difficult to define. Also, the apparent presence of dual and multiple scattering at in principle all detected energies indicates a possible contribution of nuclear losses to the observed spectrum width. This observation further justifies the necessity of employing accurate Monte-Carlo simulations including multiple scattering for the quantitative analysis of spectra. To study the impact of differences in the description of scattering, Fig. 1 presents three different MC simulations obtained by TRBS for the ZBL (green solid line) and TFM potential (blue dashed and red dash-dotted line) without any screening corrections. For the latter potential simulations with two different linear modifications of electronic stopping are shown.

To ensure proper number of incident particles in both experiment and simulation, the latter is normalized to the height of the experimental spectrum at the high energy edge $H_0$, indicated in Fig. 1. This feature of the spectra is expected to exhibit minimum influence of multiple scattering as discussed in [33] and later confirmed in the analysis presented in Fig. 2. Note, that for 25 keV He$^+$ scattered from Hf, the scattering cross section for single scattering in the employed geometry obtained with the ZBL potential exceeds the one calculated from TFM by ~10 % (not apparent due to the applied normalization). When comparing the simulations with different potentials, one can observe several clear differences. For identical electronic stopping only one of the simulations fits the width of the signal due to scattering from Hf in the experiment. For simulations with the TFM potential comparatively higher electronic stopping is needed in contrast to ZBL with a difference of ~ 3.4 % to obtain an equally precise fit of the trailing edge in the experiment. The inaccuracies in the fit to the trailing edge observed for both potentials, although to different extent, are originating from several sources. Energy loss straggling and inhomogeneous film thickness or impurities at the interface are major contributions expected. Note, however, that typically, the fit quality of the trailing edge significantly improves for higher energies, which leaves energy loss straggling which is insufficiently described in the simulations as one major source. Another factor affecting the shape of the trailing edge is the accuracy obtained by the simulations modelling the transition region from single to more extensive plural and multiple scattering yielding the spectral contribution extending to lower energies.

The intensity of the MS background changes significantly for simulations using different potentials. This difference can also be seen from the inset of Fig. 1, which shows the deviations between the depicted simulations and the experimental data. Though both simulations overestimate the background, the TFM potential gives a better overall fit of the experiment. Therefore, we perform all further simulations shown in this study with the TFM potential.

In terms of quantification, the described difficulty to define the trailing edge, leads to an intrinsic systematic uncertainty in deduced quantities, i.e. thicknesses or electronic energy loss data on a level of at least 1 % - 2 % with values of up to 5% at energies of only a few keV [34]. The observed uncertainty in simulating the multiple scattering contributions hampers additionally analysis of light species or multilayer systems, for which systematic uncertainties on the order of several percent have to be readily expected, even if a more precise fitting may be obtained.

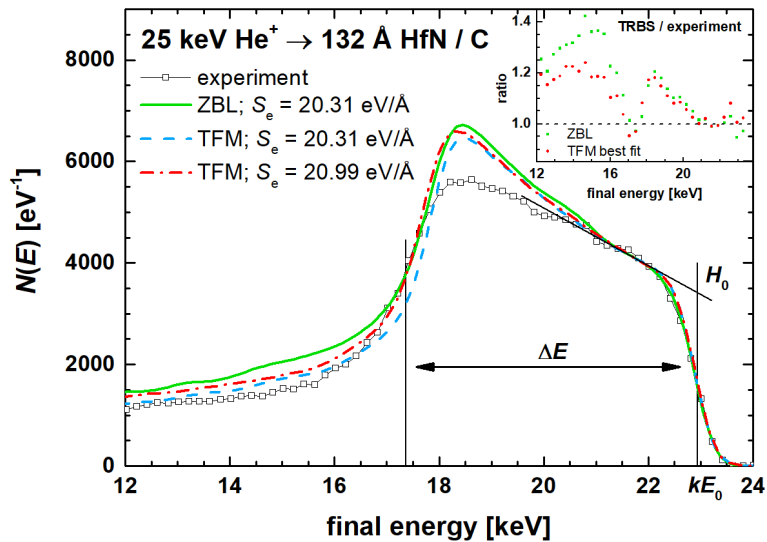

Fig. 1: Experimental energy converted ToF-MEIS data of 25 keV He$^+$ scattered from HfN (black open squares) compared to different simulations using TRBS. Simulations are plotted for both ZBL (green solid line) and TFM (blue dashed and red dash-dotted line) potentials without screening length corrections. Different electronic stopping values (also in first order equivalent to simulating different thicknesses) are required to reproduce the experimental spectrum. The inset shows the difference between simulations and experiment again for ZBL (green squares) and the best fit TFM (red circles) potential.

In Fig. 2 we present a TRBS simulation without considering any experimental resolution, again for 25 keV He$^+$ scattered from HfN. The black line represents backscattered projectiles from the whole 132 Å thick film on C. Additionally, the yield from the simulation has been plotted for ten equally thick slabs of 13.2 Å to analyze the depth dependence of the scattering yield (depicted in different colors). At the high energy edge, only projectiles scattered from the first slab contribute to the height. Also, the shape, even for a comparably low primary energy of 25 keV is fitting expectations from the single scattering model. The fact that single scattering is generally responsible for the signal at highest energies will be further corroborated by the results to be presented later in Fig. 4. For projectiles scattered from deeper layers multiple scattering increases yielding a pronounced rise in the Hf peak plateau towards the lower energy edge. As comparison the magenta line indicates the plateau height obtained with only single scattering (performed with SIMNRA [35]). Note, that the contribution to multiple scattering strongly increases with depth in a non-linear manner. In fact, the MS background can be mainly described with scattering from the last three slabs of the film (dashed dark blue line), which contribute ~ 67 % to the background. For any slab however, there is a well pronounced high-energy onset, indicating that for the highest detected energies originating from a given depth, single scattering dominates. Note, that this observation is in contrast to what is observed for scattering of heavy projectile ions and in different geometries [36].

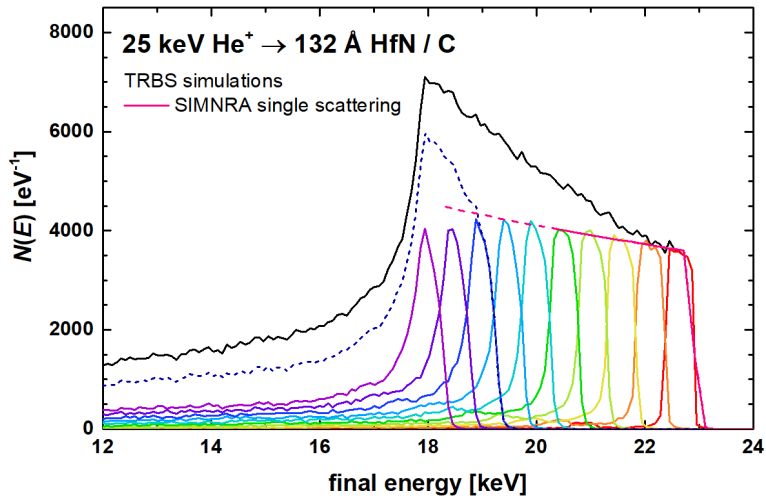

Fig. 2: TRBS simulations for 25 keV He$^+$ scattered from 132 Å of HfN on C (black line). For comparison layer resolved simulations are shown in different colors for 10 slabs á 13.2 Å. At the high energy edge, only single scattering occurs, whereas the MS background is dominated by contributions from the last three slabs (dark blue dashed line). The magenta line represents the plateau height in the single scattering model, e.g. as obtained in SIMNRA simulations.

The overestimation of the yield expected at energies below $kE_0$ apparent from Fig. 1, indicates that additional screening corrections with $c_a < 1$ are needed to properly describe the background due to multiple and plural scattering in the experiment. These results are thus an indication for a non-constant, i.e. decreasing screening length correction factor when going towards lower energies, or larger interaction distances, respectively. In Fig. 3 we show both the influence of changes in screening length as well as in electronic stopping for an energy spectrum of 30 keV He$^+$ backscattered from HfN. The red solid line corresponds to the simulation without any screening correction and a best fit of the width of the Hf peak. All depicted simulations are performed with a constant number of primary particles and we used the simulation with $c_a = 1$ for the normalization to the experiment, because only single scattering contributes to the high energy edge, as shown in Fig. 2. The interaction distances in single scattering are considered small, so that no correction in the scattering potential is necessary. The inset in Fig. 3 shows the difference between two simulations with different screening lengths ($c_a = 0.87$ vs. $c_a = 1.0$), but the same electronic stopping. The weaker screening has only a minor effect on the high energy edge, however, it deviates up to 20 % for small angle collisions.

The red solid line in Fig. 3, again overestimates the MS background, i.e. indicating a too strong potential for collisions with comparably large impact parameters. In order to reach good agreement between experiment and simulation a screening length correction of $c_a = 0.87$ is needed (blue solid line), which mainly affects the multiple scattering background and only has a minor influence on the high energy edge (see inset of Fig. 3). Note, both simulations yield roughly the same FWHM of the Hf after subtracting MS contributions. Note, also, however, that such a subtraction is not straightforward as it scales in a complex way with depth according to the results presented in Fig. 2. Consequently, this confirms the interpretation of the results from Fig. 1 indicating a systematic uncertainty on the order of 1 % - 2 % in spectrum width. To assess this, for comparison, Fig. 3 also shows simulations with a change in electronic stopping of ± 10 % as red dashed lines. As one can clearly see different stopping power values have a significant influence on the FWHM of the Hf peak.

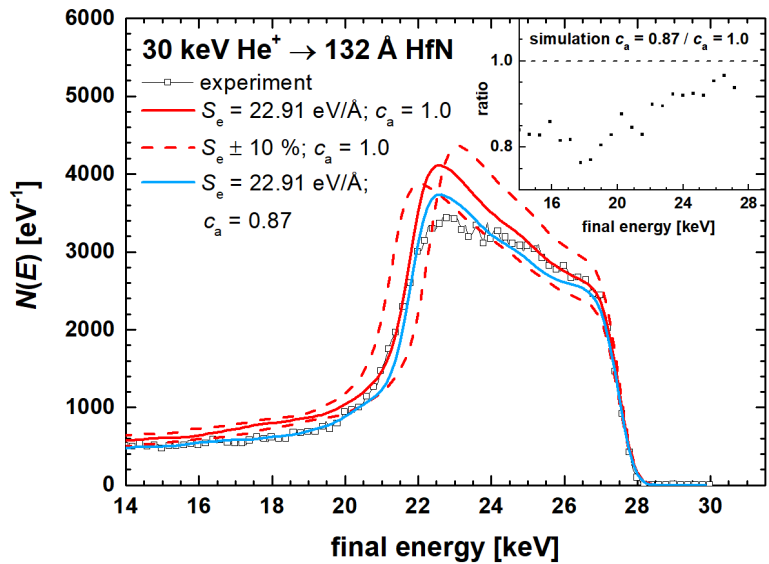

Fig. 3: Experimental energy-converted ToF-MEIS spectrum (black open squares) and TRBS simulations for 30 keV He$^+$ scattered from HfN. The red solid line shows a simulation with the best fit of the spectrum width without any screening correction. The blue solid line uses the same electronic stopping but employs the optimum

screening correction to fit the MS background. The red dashed lines correspond to a change in the electronic stopping of ± 10 %. The inset shows the difference between the red and blue simulation with and without screening corrections.

The combined results from Figs. 1 – 3, show that for quantification of sample composition in ion scattering uncertainties in the scattering potential become more important for increasing target thicknesses. However, when performing a comparison of spectra, even for compound films with finite but different thickness inaccuracies in the potential will not introduce systematically incorrect results at any given energy for which the spectra are overlapping. This intrinsic reliability can be understood, as for light ions for the signal of any depth slab, at highest energies single scattering dominates.

In the following, we have now analyzed a series of spectra obtained at different primary He energies ranging from 25-100 keV in a similar manner as for 30 keV in Fig. 3. By comparing the potential strength necessary to describe the intensity of the high energy onset $H_0$ and the multiple and plural scattering background at significantly different energy, we aimed on obtaining a more complete picture of the scattering potential. As different energies probe different impact parameters, we correlate the obtained correction factors of the screening length to specific closest interaction distances. Figure 4 depicts the screening length corrections evaluated from the low-energy background in the MEIS spectra as a function of the minimum interatomic distance for both the TFM and the ZBL potential. The distance of closest approach was obtained by the assumption that collisions with a deflection angle of around 15° represent the most relevant scattering angles for the multiple scattering background. This choice is not made arbitrarily but based on the following assumptions and observations: The inset of Fig. 4 shows a simulated energy spectrum (scattering angle $\vartheta = 155° \pm 5°$) with a relatively large cutoff angle of $\vartheta_{\min} = 15°$ (black solid line). Additionally, the simulated yields of backscattered projectiles for undergoing 1, 2, 3 as well as 4 and more collisions are plotted. Note, that the MS background in this spectrum is dominated by contributions of at least 3 collisions (~ 67 %) confirming the importance of multiple scattering over dual scattering in the background. In parallel, simulations with a series of different cut-off angles have been tested and changes in the intensity of the multiple scattering background started to exceed 10 % only for cut-off angles of 15° or higher. Specifically, a decrease in the intensity has been observed. As the collisions relevant for the low-energy background at the investigated energies in reality of course represent a range of different scattering angles, red lines in Fig. 4 represent the expected range in distance of closest approach for $10° < \vartheta < 20°$.

As the given minimum interatomic distance $r_{\min}(E, \vartheta)$ is a function of projectile energy and scattering angle, a certain $r_{\min}$ can be reached by different sets of parameters. We therefore compare the obtained estimates on the screening length corrections from small-angle scattering in the medium energy regime (energies on the left vertical axis in Fig. 4) to a typical scattering geometry in low energy ion scattering spectroscopy (LEIS) with $\vartheta = 130°$ (right vertical axis). In LEIS, screening corrections are highly relevant for quantification of single large angle scattering from surface atoms and their use is well-established. The conversion between the left and right ordinate axis is performed for the TFM potential. The according evaluation for the ZBL potential would yield a similar shift in the axis of ~ 3 %. Note, that the obtained correction factors as a function of energy fit well into the range of published values [22].

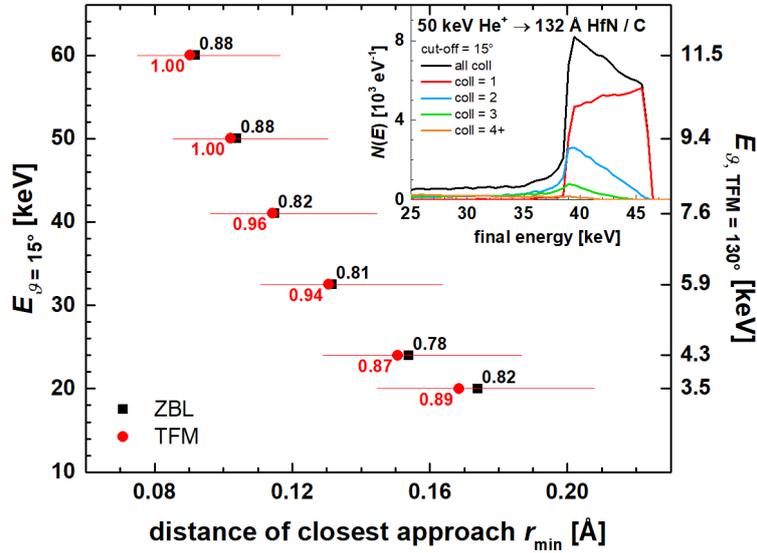

Fig. 4: Screening length corrections for both TFM and ZBL potential are depicted as a function of the parameter set $(E, \vartheta)$ and the closest approach in the scattering event. Values have been deduced from fitting the intensity of the low-energy background of the experimental spectra. The corresponding distance of closest approach in small angle scattering could be reached with different parameter sets as discussed in the introduction. For comparison, the right vertical axis shows one corresponding parameter set for a typical LEIS geometry was used. The x error bars give the uncertainty of the distance of closest approach for typical small angle scattering with $10° < \vartheta < 20°$. The inset shows a simulation with a cutoff angle of $\vartheta_{min} = 15°$; additionally the simulation is resolved for the number of undergone collisions. For details see text.

3.1.2 Subsurface layers with heavy constituents

Backscattering experiments with heavier primary ions can be of interest in thin film analysis to obtain a better mass resolution compared to standard projectiles as He. This condition applies particularly, if heavier species are located below the sample surface, and the surface layer signal can be suppressed by use of heavier ions. Vice-versa, in order to determine electronic stopping for heavier primary ions especially in low-Z materials such as e.g. TiN [14] or LiF [37] often so-called marker experiments are performed. In this case the sample consists of a thin film of the sample of interest (TiN) on top of a $\delta$-layer, where the latter consists of a high-Z element as W. Instead of analyzing the width of the signal due to scattering from Ti one evaluates the energy shift in the peak position of the $\delta$-layer, which depends on the thickness and the energy loss in TiN.

The energy converted ToF-MEIS spectrum for B$^+$ scattered from TiN/W/Si is shown in Fig. 5 as black open symbols. To obtain a good fit of the peak position of W between experiment and simulation the electronic stopping in the TiN film has to be changed accordingly with a best fit for $S_e$ = 38.19 eV/Å (red solid line). In this simulation no screening length correction is applied corresponding to the default TRBS nuclear stopping with $S_n$ = 8.11 eV/Å. The W peak can be reproduced, however the shape of the Ti peak in experiment and simulation does not coincide. We also show simulations with changes of ± 10 % in either electronic stopping or screening length as red dashed and blue dotted lines, respectively. Changes in the electronic stopping affect both shape and position of the Ti and W peak. Different screening strengths are found to have a much stronger influence on the signal at energies as corresponding to scattering from Ti, a feature which can be again assigned to the sensitivity of the multiple scattering to the specific potential strength. No significant influence on the positions of features is observed for both changes, in the potential as well as in the screening. Thus, even for heavier ions screening corrections can be neglected in energy loss evaluations from the peak position in a marker experiment. Vice versa, depth scales remain rather unaffected from uncertainties in the screened potential.

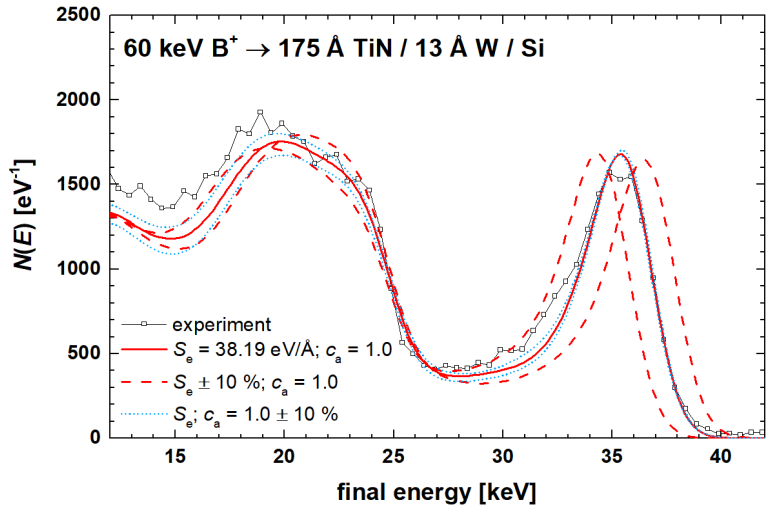

Fig. 5: Experimental and simulated spectra for B$^+$ scattered from a sample consisting of a TiN film on a W interlayer and Si substrate. The red solid line corresponds to a best fit simulation of the W peak without any screening corrections. Additionally, simulations with either changes in screening length (blue dotted line) or electronic stopping (red dashed line) are depicted.

### 3.2 Transmission simulations

As a complement to the experiments and simulations discussed above, we performed simulations in transmission geometry and for implantation depth profiles (section 3.3). While transmission experiments are less common for material analysis, they provide the most straightforward experimental approach in terms of geometry used for assessing the energy loss of energetic particles traversing matter. As the necessity of deflection is abandoned, or in contrast, even undesirable, a different impact parameter regime can be expected to be probed, in dependence of the ion energy. In energy spectra of transmitted projectiles, again the peak position of the detected particles is evaluated, which depends on the thickness as well as on the energy loss processes in the foil. In this section we investigate the observable influence of modifications in screening and electronic stopping on transmission simulations for $He^+$ and $Ne^+$ projectiles. A HfN film with a thickness of 278 Å has been used to facilitate a comparison between backscattering and transmission. A trajectory comprising of a straight line will in this case be identical in length with the shortest possible trajectory for backscattering in a single collision from the backside of the film in the geometry employed in the previous sections. Figure 6 depicts the simulated spectra for (a) $He^+$ as well as (b) and (c) $Ne^+$ transmitted through a HfN film with an acceptance angle of $\vartheta \in [0°, 2°]$. Simulations with either variation of screening length or electron stopping were performed for both projectiles. For He the obtained electronic stopping from section 3.1 serves as reference (red solid line in Fig. 6a) and the Ne simulations use the default TRBS value for stopping and no screening correction (plotted as red solid line in Fig. 6b and c). Spectra for $S_e \pm 10\%$ are plotted as red dashed lines and simulations for screening values between $c_a = 0.7$ and $c_a = 1.0$ are shown as black, green and blue lines, respectively. In this simulation 30 keV $He^+$ ions transmitted through HfN lose ~ 5.97 keV (upper graph (a)) with $S_n = 1.58$ eV/Å vs. $S_e = 22.91$ eV/Å. For $c_a = 0.9$ ($S_n = 1.65$ eV/Å) the shift in peak position is < 0.01 keV which is the applied binning in the simulation although one may expect a shift of (1 - 2) bins from the difference in nuclear stopping. In contrast, a 10 % change in electronic stopping yields the expected difference in the energy loss of ~ 0.55 keV. Therefore, small deviations in the screening length can be neglected for evaluating the energy loss from transmission measurements for light ions in the present energy regime. This finding is in accordance with expectations from the much lower magnitude of nuclear stopping and the fact that a very effective selection of trajectory is performed by the specific choice of the impact parameter [29].

The lower two panels in Fig. 6 depict simulations for 30 keV $Ne^+$ ions, where the projectiles in the reference simulation lose ~ 10.2 keV. The use of a heavier projectile leads to a significant increase in the FWHM of the transmitted peak due to more pronounced multiple and plural scattering contributions. Note also, that nuclear stopping is dominant over electronic stopping in the present energy regime ($S_n = 49.23$ eV/Å and $S_e = 28.09$ eV/Å). A comparison to simulations with ± 10 % in electronic stopping is plotted in panel (b) yielding a peak shift of ~ 0.8 keV for both cases. In the lower graph (c) variations in the screening length are depicted with a shift of the maximum of roughly 0.6 keV for a change in the screening length correction by 0.1. Note, a comparison between simulations with TFM and ZBL potential yields already a shift of the peak position of ~ 0.2 keV (the energy spectrum obtained with ZBL potential is not depicted in Fig. 6c). For both projectiles, He and Ne, a more drastic effect is observed in a change of the transmitted intensity per impinging ion, due to modification of the average angular deflection.

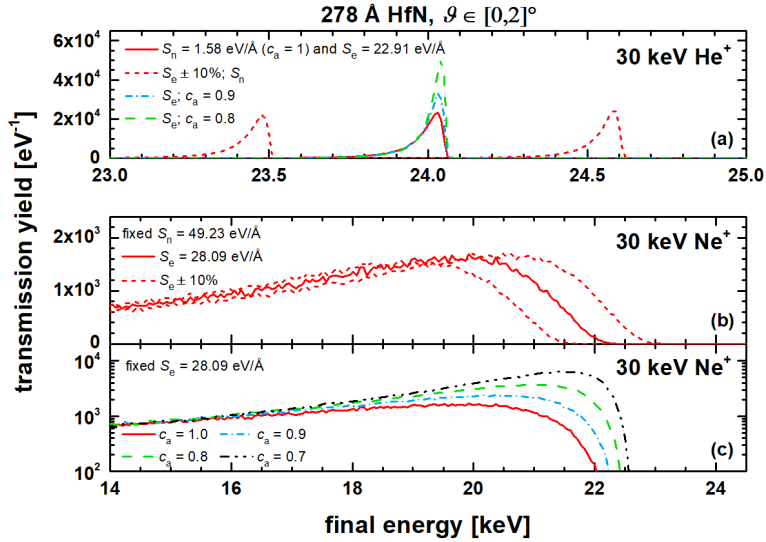

Fig. **Error! Unknown switch argument.**: Simulated spectra for the transmitted ion yield of (a) He$^+$ and (b), (c) Ne$^+$ through HfN. In all three panels the red solid line represents a simulation with the default SRIM stopping used in the TRBS code and no screening corrections. Variations in the electronic stopping are depicted as dashed red lines and changes in the screening by the blue, green and black lines.

3.4 Ion implantation

For the depth profiles of implanted ions (not depicted in this contribution), we performed simulations with the same parameters as for the transmission geometry concerning change in electronic stopping and screening length correction. The HfN thickness was modified, so that all primary projectiles are stopped in the film. As a result, the He distribution in the film is found to change in a similar way for 10 % change in either electronic stopping or screening length. In all cases increasing potential strength and electronic stopping decreased the position of the maximum in the implantation profile by around 5 % of the range and vice versa. The non-linear scaling can be explained by the influence of scattering on the ratio of path length to projected range which was investigated here. For Ne implantations, however, the same modification of the screening length yields a variation in the position of the maximum by roughly 10 %, whereas changes in electronic stopping only have a minor effect on the range of ions. The increased relative weight of corrections to the potential in implantations can be understood on the basis of trajectory selection being important in both backscattering and transmission experiments as shown throughout this work. As only specific trajectories with low scattering probability are probed the influence of the potential is minimized, with benefits for quantitative analysis.

## 4. Summary and conclusions

In this contribution, we evaluated the influence of uncertainties in the screened Coulomb potentials on the shape of MEIS energy spectra, affecting either quantification of sample composition or evaluation of electronic stopping, in backscattering as well as transmission geometry.

For He projectiles, a comparison between energy spectra from backscattering experiments and MC simulations showed significant deviations in the MS background for primary energies below 100 keV. In general, it was found, that the use of the TFM potential yields a better overall fit between the spectra shape of experiment and simulation compared to the ZBL potential. However, both potentials overestimate the multiple scattering background for lower primary energies. Small screening corrections as employed in this contribution have only a minor influence on the evaluation of the spectrum width (1 % - 2 %); however, a systematic difference between the TFM and the ZBL potential towards lower energies was found with a resulting difference in electronic energy loss evaluations of ~ 3 % for the lowest investigated energy. These deviations in the scattering potential, which depend on projectile-target system as well as sample thickness, directly influence quantification with low- and medium-energy ion scattering.

For heavier primary ions as B electronic stopping is still larger than nuclear stopping in the investigated energy regime, however the nuclear losses for heavier projectiles are significantly larger compared to He. These measurements were performed in double transmission geometry, where the position of the $\delta$-layer peak is evaluated. We did not see an influence of either employed potentials or screening corrections in the peak position of the W layer that goes beyond observations made for light projectiles.

In transmission geometry simulations for two systems have been performed: (a) electronic stopping (He projectiles) or (b) nuclear stopping (Ne projectiles) dominating the total energy loss in the film. When electronic stopping is dominating, the influence of the employed potential as well as additional screening corrections on the peak position is negligible. However, for heavier projectiles like Ne, where nuclear stopping dominates, the influence of screening corrections is in the same order of magnitude as changes in electronic stopping.

Altogether the results indicate that analytical approaches based on spectrometry of primary ions strongly benefit from effective trajectory selection minimizing the influence of uncertainties in the potential significantly beyond its expected impact on the average ion trajectory in the target material.

**Acknowledgment:**

This work was supported by the Swedish Foundation for Strategic Research (SSF) (contract #RIF14-0053) as well as the VR-RFI (contract #821-2012-5144 and #2017-00646_9). Additionally, BB is grateful to the Wilhelm-Macke Foundation at the JKU for partial support of her stay at UU.

**Bibliography**

[1]  J. Hämäläinen, M. Mattinen, K. Mizohata, K. Meinander, M. Vehkamäki, J. Räisänen, M. Ritala, M. Leskelä, Atomic Layer Deposition of Rhenium Disulfide, Adv. Mater. 30 (2018) 1703622. doi:10.1002/adma.201703622.

[2]  E. Stevens, Y. Tomczak, B.T. Chan, E. Altamirano Sanchez, G.N. Parsons, A. Delabie, Area-Selective Atomic Layer Deposition of TiN, $TiO_2$, and $HfO_2$ on Silicon Nitride with inhibition on Amorphous Carbon, Chem. Mater. 30 (2018) 3223–3232. doi:10.1021/acs.chemmater.8b00017.

[3]  Y. Zhao, L. V. Goncharova, A. Lushington, Q. Sun, H. Yadegari, B. Wang, W. Xiao, R. Li, X. Sun, Superior Stable and Long Life Sodium Metal Anodes Achieved by Atomic Layer Deposition, Adv. Mater. 29 (2017) 1606663. doi:10.1002/adma.201606663.

[4]  J.R. Tesmer, M.A. Nastasi, eds., Handbook of modern ion beam materials analysis, Materials Research Society, Pittsburgh, Pennsylvania, 1995.

[5]  H. Li, L. Daukiya, S. Haldar, A. Lindblad, B. Sanyal, O. Eriksson, D. Aubel, S. Hajjar-Garreau, L. Simon, K. Leifer, Site-selective local fluorination of graphene induced by focused ion beam irradiation, Sci. Rep. 6 (2016) 19719. doi:10.1038/srep19719.

[6]  P. Willke, J.A. Amani, A. Sinterhauf, S. Thakur, T. Kotzott, T. Druga, S. Weikert, K. Maiti, H. Hofsäss, M. Wenderoth, Doping of Graphene by Low-Energy Ion Beam Implantation: Structural, Electronic, and Transport Properties, Nano Lett. 15 (2015) 5110–5115. doi:10.1021/acs.nanolett.5b01280.

[7]  E. Rutherford, LXXIX. *The scattering of α and β particles by matter and the structure of the atom*, Philos. Mag. 21 (1911) 669.

[8]  J.F. Ziegler, J.P. Biersack, The Stopping and Range of Ions in Matter, Springer US, 1985. doi:10.1007/978-1-4615-8103-1_3.

[9]  Y. Kido, T. Nishimura, Y. Hoshino, H. Namba, Surface structures of SrTiO3(0 0 1) and Ni/SrTiO3(0 0 1) studied by medium-energy ion scattering and SR-photoelectron spectroscopy, Nucl. Instruments Methods Phys. Res. Sect. B Beam Interact. with Mater. Atoms. 161–163 (2000) 371–376. doi:10.1016/S0168-583X(99)00715-6.

[10] M. Riva, M. Kubicek, X. Hao, G. Franceschi, S. Gerhold, M. Schmid, H. Hutter, J. Fleig, C. Franchini, B. Yildiz, U. Diebold, Influence of surface atomic structure demonstrated on oxygen incorporation mechanism at a model perovskite oxide, Nat. Commun. 9 (2018) 3710. doi:10.1038/s41467-018-05685-5.

[11] H.H. Brongersma, M. Draxler, M. de Ridder, P. Bauer, Surface composition analysis by low-

energy ion scattering, Surf. Sci. Rep. 62 (2007) 63–109. doi:10.1016/j.surfrep.2006.12.002.

[12] A.A. Zameshin, A.E. Yakshin, J.M. Sturm, H.H. Brongerma, F. Bijkerk, Double matrix effect in Low Energy Ion Scattering from La surfaces, Appl. Surf. Sci. 440 (2018) 570–579. doi:10.1016/j.apsusc.2018.01.174.

[13] P. Kürnsteiner, R. Steinberger, D. Primetzhofer, D. Goebl, T. Wagner, Z. Druckmüllerova, P. Zeppenfeld, P. Bauer, Matrix effects in the neutralization of He ions at a metal surface containing oxygen, Surf. Sci. 609 (2013) 167–171. doi:10.1016/j.susc.2012.12.003.

[14] M.A. Sortica, V. Paneta, B. Bruckner, S. Lohmann, T. Nyberg, P. Bauer, D. Primetzhofer, On the Z1-dependence of electronic stopping in TiN, Sci. Rep. 9 (2019) 176. doi:10.1038/s41598-018-36765-7.

[15] E.E. Quashie, A.A. Correa, Electronic stopping power of protons and alpha particles in nickel, Phys. Rev. B. 98 (2018) 235122. doi:10.1103/PhysRevB.98.235122.

[16] H.H. Andersen, F. Besenbacher, P. Loftager, W. Möller, Large-angle scattering of light ions in the weakly screened Rutherford region, Phys. Rev. A. 21 (1980) 1891–1901. doi:10.1103/PhysRevA.21.1891.

[17] G. Molière, Theorie der {Streuung} schneller geladener {Teilchen} {I}. {Einzelstreuung} am abgeschirmten {Coulomb-Feld}, Zeitschrift Naturforsch. Tl. A. 2 (1947) 133.

[18] J.F. Ziegler, J.P. Biersack, U. Littmark, The Stopping and Range of Ions in Solids, Pergamon Press. (1985).

[19] O.B. Firsov, Calculation of the interaction potential of atoms, Sov. Phys. JETP. 6 (1958).

[20] D.J. O'connor, J.P. Biersack, Comparison of theoretical and empirical interatomic potentials, Nucl. Instruments Methods Phys. Res. Sect. B Beam Interact. with Mater. Atoms. 15 (1986) 14–19. doi:10.1016/0168-583X(86)90243-0.

[21] W. Takeuchi, Evaluation of screening length corrections for interaction potentials in impact-collision ion scattering spectroscopy, Nucl. Instruments Methods Phys. Res. Sect. B Beam Interact. with Mater. Atoms. 313 (2013) 33–39. doi:10.1016/J.NIMB.2013.08.003.

[22] D. Primetzhofer, S.N. Markin, M. Draxler, R. Beikler, E. Taglauer, P. Bauer, Strength of the interatomic potential derived from angular scans in {LEIS}, Surf. Sci. 602 (2008) 2921–2926. doi:10.1016/j.susc.2008.07.030.

[23] M. Draxler, M. Walker, C.F. McConville, Determination of a correction factor for the interaction potential of He+ ions backscattered from a Cu(1 0 0) surface, Nucl. Instruments Methods Phys. Res. Sect. B Beam Interact. with Mater. Atoms. 249 (2006) 812–815.

doi:10.1016/J.NIMB.2006.03.177.

[24] R. Souda, M. Aono, C. Oshima, S. Otani, Y. Ishizawa, Multiple scattering of low-energy rare-gas ions at solid surfaces, Surf. Sci. 176 (1986) 657–668. doi:10.1016/0039-6028(86)90064-6.

[25] A.N. Zinoviev, Electron screening of the {Coulomb} potential at small internuclear distances, Nucl. Instruments Methods Phys. Res. Sect. B Beam Interact. with Mater. Atoms. 354 (2015) 308–312. doi:10.1016/j.nimb.2014.11.085.

[26] A.N. Zinoviev, Potentials of the Interaction of Atomic Particles at Large, Medium, and Small Collision Energies, J. Surf. Investig. X-Ray, Synchrotron Neutron Tech. 12 (2018) 554–557. doi:10.1134/S1027451018030382.

[27] F. Matias, R.C. Fadanelli, P.L. Grande, N.E. Koval, R.D. Muiño, A.G. Borisov, N.R. Arista, G. Schiwietz, Ground- and excited-state scattering potentials for the stopping of protons in an electron gas, J. Phys. B At. Mol. Opt. Phys. 50 (2017) 185201. doi:10.1088/1361-6455/aa843d.

[28] P. Sigmund, A. Schinner, Is electronic stopping of ions velocity-proportional in the velocity-proportional regime?, Nucl. Instruments Methods Phys. Res. Sect. B Beam Interact. with Mater. Atoms. 440 (2019) 41–47. doi:10.1016/J.NIMB.2018.10.031.

[29] S.R. Naqvi, G. Possnert, D. Primetzhofer, Energy loss of slow Ne ions in Pt and Ag from TOF-MEIS and Monte-Carlo simulations, Nucl. Instruments Methods Phys. Res. Sect. B Beam Interact. with Mater. Atoms. 371 (2016) 76–80. doi:http://dx.doi.org/10.1016/j.nimb.2015.09.048.

[30] M.K. Linnarsson, A. Hallén, J. Åström, D. Primetzhofer, S. Legendre, G. Possnert, New beam line for time-of-flight medium energy ion scattering with large area position sensitive detector, Rev. Sci. Instrum. 83 (2012) 095107. doi:10.1063/1.4750195.

[31] D. Primetzhofer, E. Dentoni Litta, A. Hallén, M.K. Linnarsson, G. Possnert, Ultra-thin film and interface analysis of high-k dielectric materials employing Time-Of-Flight Medium Energy Ion Scattering (TOF-MEIS), Nucl. Instruments Methods Phys. Res. Sect. B Beam Interact. with Mater. Atoms. 332 (2014) 212–215. doi:10.1016/j.nimb.2014.02.063.

[32] J.P. Biersack, E. Steinbauer, P. Bauer, A particularly fast TRIM version for ion backscattering and high energy ion implantation, Nucl. Inst. Methods Phys. Res. B. 61 (1991) 77. doi:10.1016/0168-583X(91)95564-T.

[33] D. Roth, D. Goebl, D. Primetzhofer, P. Bauer, A procedure to determine electronic energy loss from relative measurements with TOF-LEIS, Nucl. Instruments Methods Phys. Res. Sect. B Beam Interact. with Mater. Atoms. 317 (2013) 61–65. doi:10.1016/j.nimb.2012.12.094.

[34] S.N. Markin, D. Primetzhofer, S. Prusa, M. Brunmayr, G. Kowarik, F. Aumayr, P. Bauer,

Electronic interaction of very slow light ions in Au: Electronic stopping and electron emission, Phys. Rev. B. 78 (2008) 195122. doi:10.1103/PhysRevB.78.195122.

[35] M. Mayer, SIMNRA, a simulation program for the analysis of NRA, RBS and ERDA, in: AIP Conf. Proc., AIP, 1999: pp. 541–544. doi:10.1063/1.59188.

[36] K. Kantre, V. Paneta, D. Primetzhofer, Investigation of the energy loss of I in Au at energies below the Bragg peak, Nucl. Instruments Methods Phys. Res. Sect. B Beam Interact. with Mater. Atoms. (2019). doi:10.1016/J.NIMB.2018.10.034.

[37] S.N. Markin, D. Primetzhofer, P. Bauer, Vanishing electronic energy loss of very slow light ions in insulators with large band gaps, Phys. Rev. Lett. 103 (2009). doi:10.1103/PhysRevLett.103.113201.